\pdfoutput=1
\documentclass[prl,aps,twocolumn,preprintnumbers,nofootinbib,superscriptaddress,showpacs,floatfix]{revtex4}
\usepackage{amsmath,amssymb,amsbsy}
\usepackage{graphicx}
\usepackage{epsfig}
\usepackage{color}

\newcommand{\Delstar}{\ensuremath{\Delta^{\raise0.18ex\hbox{${\scriptstyle *}$}}}}
\def\gtwid{{\,\raise.35ex\hbox{$>$\kern-.75em\lower1ex\hbox{$\sim$}}\,}}
\def\ltwid{{\,\raise.35ex\hbox{$<$\kern-.75em\lower1ex\hbox{$\sim$}}\,}}
\def\leftvec{{\raise1.5ex\hbox{$\leftarrow$}\kern-1.00em}}
\def\rightvec{{\raise1.5ex\hbox{$\rightarrow$}\kern-1.00em}}
\def\half{{\scriptstyle \raise.2ex\hbox{${1\over2}$}}}
\def\threehalves{{\scriptstyle \raise.15ex\hbox{${3\over2}$}}}
\def\third{{\scriptstyle \raise.15ex\hbox{${1\over3}$}}}
\def\third{{\scriptstyle \raise.15ex\hbox{${1\over3}$}}}
\def\twothirds{{\scriptstyle \raise.15ex\hbox{${2\over3}$}}}
\def\fourth{{\scriptstyle \raise.15ex\hbox{${1\over4}$}}}

\newcommand*{\bea}{\begin{eqnarray}}
\newcommand*{\eea}{\end{eqnarray}}
\newcommand*{\be}{\begin{equation}}
\newcommand*{\ee}{\end{equation}}

\newcommand*{\CPT}{\raise0.4ex\hbox{$\chi$}PT}
\newcommand*{\chpt}{\raise0.4ex\hbox{$\chi$}PT}
\newcommand*{\schpt}{S\raise0.4ex\hbox{$\chi$}PT}

\def\eqref#1{{(\ref{#1})}}

\def\hat{\widehat}

\def\bea{\begin{eqnarray}}
\def\eea{\end{eqnarray}}

\def\beq{\begin{equation}}
\def\eeq{\end{equation}}

\def\spose#1{\hbox to 0pt{#1\hss}}
\def\ltapprox{\mathrel{\spose{\lower 3pt\hbox{$\mathchar"218$}}
 \raise 2.0pt\hbox{$\mathchar"13C$}}}
\def\gtapprox{\mathrel{\spose{\lower 3pt\hbox{$\mathchar"218$}}
 \raise 2.0pt\hbox{$\mathchar"13E$}}}
\def\inapprox{\mathrel{\spose{\lower 3pt\hbox{$\mathchar"218$}}
 \raise 2.0pt\hbox{$\mathchar"232$}}}

\begin{document}

\preprint{IU-HET-551}


\title{Possible evidence for the breakdown of the CKM-paradigm of CP-violation}

\author{E.\ Lunghi}
\affiliation{Physics Department, Indiana University, Bloomington, IN 47405}

\author{Amarjit Soni}
\affiliation{Physics Department, Brookhaven National Laboratory, Upton, NY 11973, USA }

\begin{abstract}
Using primarily experimental inputs for S($B_d \to \psi K_s$), $\Delta M_{B_s}$, $\Delta M_{B_d}$, {\rm BR} ($B \to \tau \nu$) and $\epsilon_K$ along with necessary inputs from the lattice, we find that the measured value of $\sin (2 \beta)$ is smaller than expectations of the Standard Model by as much as 3.3 $\sigma$, and also that the measured value of the ${\rm BR} (B \to \tau \nu)$ seems to be less than the predicted value by about 2.8 $\sigma$. However, through a critical study we show that most likely the dominant source of these deviations is in $B_{d(s)}$ mixings and in $\sin (2 \beta)$ and less so in $B \to \tau \nu$, and also that the bulk of the problem persists even if input from $\epsilon_K$ is not used. The fact that kaon mixing and $\epsilon_K$ are not the dominant source of the deviation from the Standard Model has the very important consequence that model independent considerations imply that the scale of the relevant new CP-violating physics is below $O(2 \; {\rm TeV})$, and possibly even a few hundred GeVs, thus suggesting that direct signals of the new particle(s) may well be accessible in collider experiments at the LHC and perhaps even at the Tevatron.

\end{abstract}
\pacs{12.15.Hh,11.30.Hv}
\maketitle

\noindent {\bf Introduction.} 
It has been clear for quite sometime now that the CKM-paradigm~\cite{ckm} of the Standard Model (SM) provides a quantitative description of the observed CP violation, simultaneously in the B-system as well as in the K-system with a single CP-odd phase, to an accuracy of about 20\%~\cite{Nir02}. This major milestone in our understanding of CP violation was, of course, achieved in large part due to the spectacular performance of the two asymmetric B-factories. However, it should be recognized that input from the lattice for various weak matrix elements has also played a crucial role; in particular, the input from the lattice for the precise value of the hadronic parameter (``$B_K$") characterizing the kaon-mixing amplitude is essential to demonstrate a quantitative understanding of the indirect CP-violation in $K_L \to \pi \pi$, {\it i.e.} $\epsilon_K \approx 10^{-3}$. While the success of the CKM picture is very impressive, the flip side is that an accuracy of $O(20\%)$ leaves open the possibility of quite sizable new physics contributions. Indeed, in the past few years as better data and better theoretical calculations became available some tensions have emerged~\cite{Lunghi:2007ak,Lunghi:2008aa, Bona:2009cj,Lunghi:2009sm,Lunghi:2009ke,Lenz:2010gu}. It is clearly important to scrutinize these tensions carefully to see if they persist and if so what they imply for the possible existence of new physics.
 
In this paper we show that use of the latest experimental inputs along with a careful use of the latest lattice results leads to a rather strong case in favor of a possible failure of the CKM picture for a sizable contribution due to beyond the Standard Model sources of CP violation that in $\sin 2 \beta$ could be around 15-25\%. Clearly if this result stands further scrutiny it would have widespread and significant repercussions for experiments at the intensity as well as the high energy frontier and, of course, also for our theoretical understanding. We will also show that we are able to isolate the presence of new physics to $\Delta B=2$ mixing amplitudes with a possible sub-dominant effect in kaon-mixing. In particular our analysis indicates  that the data does not seem to provide a consistent interpretation for the presence of large new physics contribution to the tree amplitude for $B\to \tau \nu$. \\

\noindent {\bf Choice of lattice inputs.}
Key inputs from experiment and from the lattice needed for our analysis are shown in Table~\ref{tab:utinputs}. Below we briefly remark on a few of the items here that deserve special mention:

\noindent $\bullet$   With regard to lattice inputs we want to emphasize that quantities used here have been carefully chosen and  are extensively studied on the lattice for many years. Results that we use are obtained in full QCD with $N_F= 2 + 1$ simulations, are in the continuum limit, are fairly robust and emerge from the works of more than one collaboration and in most cases by many collaborations. 

\noindent $\bullet$   For using $B\to\tau\nu$ in the context of  our fits one clearly needs $f_B$. Although this can be, and is directly calculated on the lattice and results are known (see Table~\ref{tab:utinputs}), we will not use this as an input. Rather, our strategy is to extract the ``predicted" or fitted value of $f_B$ in the context of a given hypothesis so that comparison with the value directly  determined from lattice calculations then can serve as a useful test of the viability of that particular hypothesis. As we will show, this proves to be a rather useful criteria in checking the internal consistency of the hypotheses. This strategy will  help us to isolate the main source of the deviation between experiment and the SM and thereby it may also serve as a useful way to uncover the nature of the underlying new physics. For completeness let us mention that $f_B$ needed for $B \to \tau \nu$ in conjunction with our analysis is deduced indirectly by using the SU(3) breaking ratio, $\xi = f_{B_s} \hat B_s^{1/2} / f_{B} \hat B_d^{1/2}$, $f_{B_s} \hat B_s^{1/2}$ and $\hat B_d$.

\noindent $\bullet$   Regarding calculations of $\hat B_K$ on the lattice, it is useful to note that in the past 3 years a dramatic reduction in errors  has been achieved and by now a number of independent calculations with errors $\ltwid 5\%$ and with consistent central values have been obtained  using $N_f = 2 + 1$~\cite{Gamiz:2006sq,RBC-UKQCD,Aubin:2009jh,Kim:2009te,Bae:2010ki} as well as $N_f = 2$~\cite{Aoki:2008ss} dynamical simulations~\cite{Laiho:2010conf}. Again, to be conservative we only use weighted average of two results that are both in full QCD, use different fermion discretizations (domain-wall and staggered) and that have no correlations between them~\cite{RBC-UKQCD,Bae:2010ki}. 

\noindent $\bullet$   Regarding $V_{ub}$, since inclusive and exclusive methods differ appreciably, it should be clear that it is very difficult to draw  reliable conclusions by using this quantity; we will therefore make very limited and peripheral use of $V_{ub}$ only. For $|V_{ub}|_{\rm excl}$, to be conservative, we take the smaller of the two errors between the FNAL/MILC~\cite{Dalgic:2006dt} and HPQCD~\cite{Bailey:2008wp} collaborations rather than taking the weighted average . Also, we add a $10\%$ uncertainty to the inclusive determination of $V_{ub}$ in order to reflect intrinsic uncertainties of the theoretical model adopted. The exclusive and inclusive determinations still differ at the $1.8\sigma$ level; see Table \ref{tab:utinputs}.

\noindent $\bullet$   In a nutshell, we want to emphasize that we believe that the inputs used from the lattice are robust and reliable and therefore the implications resulting from their application should be taken seriously. \\

\begin{table}[t]
\begin{center} 
\begin{tabular}{ll}
\toprule
$\left| V_{cb} \right|_{\rm excl} = (39.0 \pm 1.2) 10^{-3}$& 
$\eta_1 = 1.51 \pm 0.24$~\cite{Herrlich:1993yv} \cr
$\left| V_{cb} \right|_{\rm incl} = (41.31 \pm 0.76) 10^{-3}$&  
$\eta_2 = 0.5765 \pm 0.0065$~\cite{Buras:1990fn} \cr
$\left| V_{cb} \right|_{\rm tot} = (40.43 \pm 0.86) 10^{-3}$  & 
$\eta_3 = 0.494 \pm 0.046$~\cite{Herrlich:1995hh,Brod:2010mj} \cr
$\left| V_{ub} \right|_{\rm excl} = (29.7 \pm 3.1) 10^{-4}$  &
$\eta_B = 0.551 \pm 0.007$~\cite{Buchalla:1996vs}\cr
$\left| V_{ub} \right|_{\rm incl} =  (40.1 \pm 2.7 \pm 4.0) 10^{-4} $    & 
$\xi = 1.23 \pm 0.04$~\cite{Gamiz:2009ku,Gamiz:2010private}\cr
$\left| V_{ub} \right|_{\rm tot} = (32.7 \pm 4.7) 10^{-4}$  & 
$\lambda = 0.2255  \pm 0.0007$ \cr
$\Delta m_{B_d} = (0.507 \pm 0.005)\; {\rm ps}^{-1}$ & 
$\alpha = (89.5 \pm 4.3)^{\rm o}$\cr
$\Delta m_{B_s} = (17.77 \pm 0.12 )\;  {\rm ps}^{-1}$ & 
$\kappa_\varepsilon = 0.94 \pm 0.02$~\cite{Buras:2008nn,Laiho:2009eu,Buras:2010pza} \cr
$S_{\psi K_S} = 0.668 \pm 0.023$~\cite{MK_10}&
$\gamma = (74 \pm 11)^{\rm o}$ \cr
$m_c(m_c) = (1.268 \pm 0.009 ) \; {\rm GeV}$ &
$\hat B_K = 0.740 \pm 0.025$ \cr
$m_{t, pole} = (172.4 \pm 1.2) \; {\rm GeV}$ & 
$f_K = (155.8 \pm 1.7) \; {\rm MeV}$ \cr
$f_{B_s} \sqrt{\hat B_s}  =  (276 \pm 19) \; {\rm MeV}$~\cite{Gamiz:2009ku} &
 $\varepsilon_K = (2.229 \pm 0.012 ) 10^{-3}$\cr
$f_B = (208 \pm 8) \; {\rm MeV}$~\cite{Gamiz:2009ku,Gamiz:2010private}~\footnote{Our value of $f_B$ reflects the change in the overall scale ($r_1$) recently adopted by the Fermilab/MILC and HPQCD collaborations~\cite{claude}} &
 $\hat B_d = 1.26 \pm 0.10$~\cite{Gamiz:2009ku,Gamiz:2010private}\cr
\multicolumn{2}{l}{${\cal B}_{B\to \tau\nu} = (1.68\pm 0.31) \times 10^{-4}$~\cite{Ikado:2006un,BaBar:2010rt,Hara:2010dk}} \cr
\botrule
\end{tabular}
\caption{Inputs used in the fit. References to the original experimental and theoretical papers and the description of the averaging procedure can be found in Ref.~\cite{Laiho:2009eu}. Statistical and systematic errors are combined in quadrature. We adopt the averages of Ref.~\cite{Laiho:2009eu} for all quantities with the exception of $|V_{ub}|$, $\xi$, $f_{B_s} \hat B_s^{1/2}$, $\hat B_K$ and $f_B$ (see text). \label{tab:utinputs}}
\end{center}
\end{table}

\begin{figure}[t]
\begin{center}
\includegraphics[width=0.95 \linewidth]{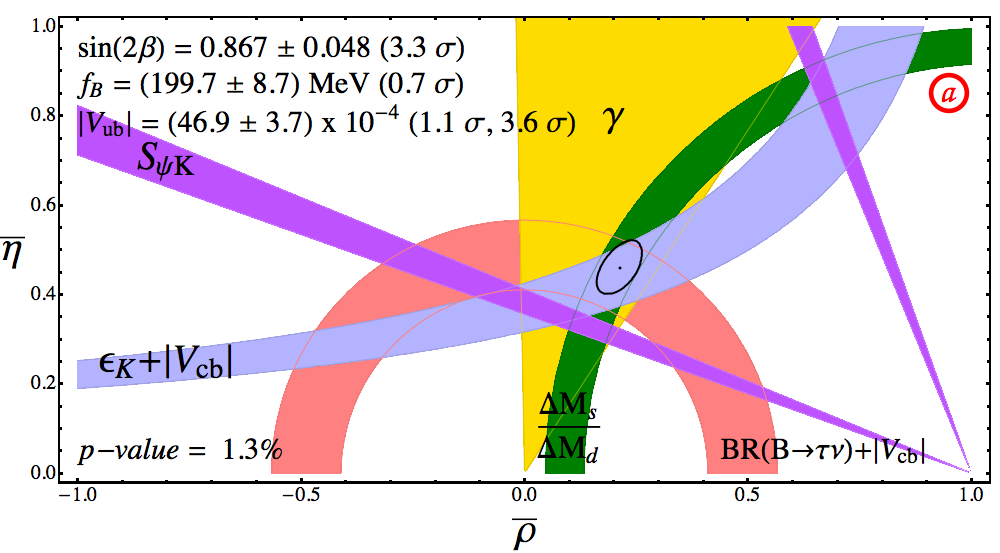}
\includegraphics[width=0.95 \linewidth]{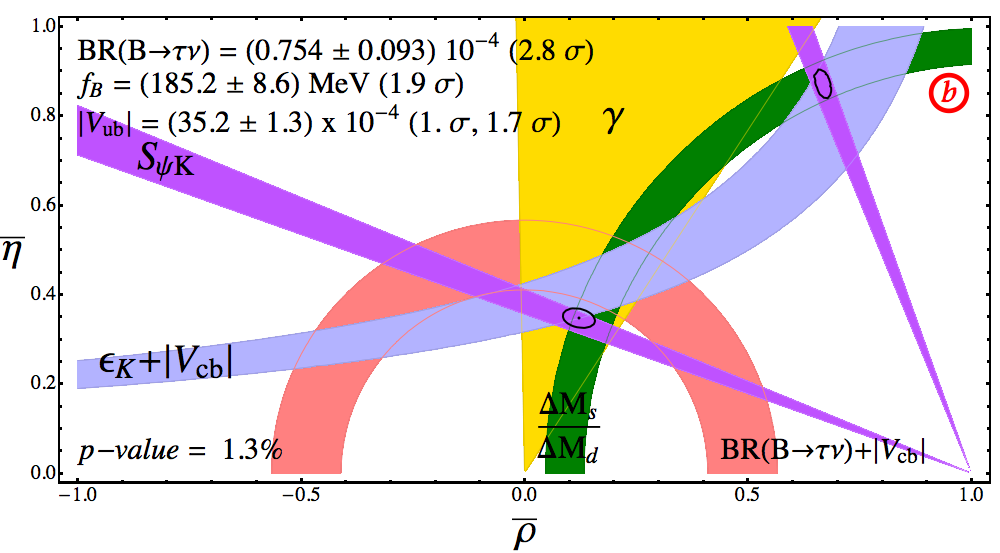}
\caption{Unitarity triangle fit. In the top panel, the contour and the fit predictions for $\sin (2\beta)$, $f_B$ and $|V_{ub}|$ are obtained using $V_{cb}$, $\varepsilon_K$, $\gamma$, $\Delta M_{B_{d,s}}$ and $B\to\tau\nu$. In the bottom panel, the contour and the fit predictions for ${\rm BR} (B\to \tau\nu)$, $f_B$ and $|V_{ub}|$ are obtained using $V_{cb}$, $\varepsilon_K$, $\gamma$, $\Delta M_{B_{d,s}}$ and $\sin(2\beta)$. 
\label{fig:utfit}}
\end{center}
\end{figure}

\noindent {\bf Result of the Fit.}
The results of the fit are shown in Fig.~\ref{fig:utfit}. In Fig.~\ref{fig:utfit}a we use as inputs from experiments, $\epsilon_K$, $\Delta M_s/\Delta M_d$, $\gamma$ and ${\rm BR} (B \to \tau \nu)$~\cite{no_alpha} and from the lattice, $\hat B_K$, $\xi$, $f_{B_s} \hat B_s^1/2$ and $\hat B_d$ (but not $f_B$) and we extract the fitted value of $\sin 2 \beta$ and of $f_B$. We obtain:
\begin{equation}
\sin(2\beta)^{\rm fit} = 0.867 \pm 0.048 \; , \label{sin2betafit}
\end{equation}
which is about 3.3 $\sigma$ away from the experimentally measured value of $0.668 \pm 0.023$. This is the key result of this paper providing a strong indication that the CKM description of the observed CP violation is breaking down~\cite{no_gamma}

Clearly if this result is true it would be rather significant; therefore we want to carefully check how robust it is and also try to isolate which of the relevant physical process or processes may be seeing the effect of new physics.

An important clue is provided by extracting the fitted value of $f_B$ along with the predicted value of $\sin (2 \beta)$ given above, we find:
\beq
f_B^{\rm fit} = (199.7 \pm 8.7)\; {\rm MeV} \; . \label{fbsin2beta}
\eeq
This ``predicted" value is in very good agreement with the one obtained by direct lattice calculation, $f_B = (208 \pm 8)$ MeV. This is a useful consistency check signifying that the SM description of the inputs used, especially of $B \to \tau \nu$, is working fairly well and that it is unlikely that the $B\to \tau\nu$ tree amplitude is receiving large contributions from new physics. Most likely the dominant effect of new physics is in fact in $\sin (2 \beta)$. Note also that the fitted value of $\left| V_{ub} \right| = (46.9 \pm 3.7) \times 10^{-4}$ is quite consistent with that determined by the inclusive method and deviates significantly ($\approx 3.6~\sigma$) from that obtained by the exclusive approach. Later we will
comment on this regarding especially the issue of ${\rm BR}(B \to \tau \nu)$).

In order to  further scrutinize the tentative conclusion reached above, we next present an alternate scenario depicted in Fig.~\ref{fig:utfit}b. Here, we make one important change in the inputs used. Instead of using the measured value of ${\rm BR} (B \to \tau \nu)$ we now use as input the measured value of $\sin (2 \beta)$ from the ``gold-plated" $B_d \to \psi K_s$ mode. Again, this fit yields two important predictions:
\begin{align}
{\rm BR} (B\to \tau\nu)^{\rm fit} &=(0.754\pm0.093) \times 10^{-4} \; , \label{btnfit} \\
f_B^{\rm fit} &= (185.2 \pm 8.6)\; {\rm MeV} \; . \label{fbbtn}
\end{align}
Eq.~(\ref{btnfit}) deviates by $2.8\sigma$ from the experimental measurement, as can also be gleaned from an inspection of Fig.~\ref{fig:utfit}b. It is particularly interesting that also the fit prediction for $f_B$ now deviates by about $1.9\sigma$ from the direct lattice determination given in Table~\ref{tab:utinputs}. We believe this provides additional support that the measured value of $\sin (2 \beta)$ being used here as a key input is not consistent with the SM and in fact is receiving appreciable contributions from new physics.

This leads us to conclude that while the presence of some sub-dominant contribution of new physics in $B\to\tau\nu$ is possible, a large contribution of new physics in there is not able to explain, in a consistent fashion, the tension we are observing in the unitarity triangle fit. 

\begin{figure}
\begin{center}
\includegraphics[width=1 \linewidth]{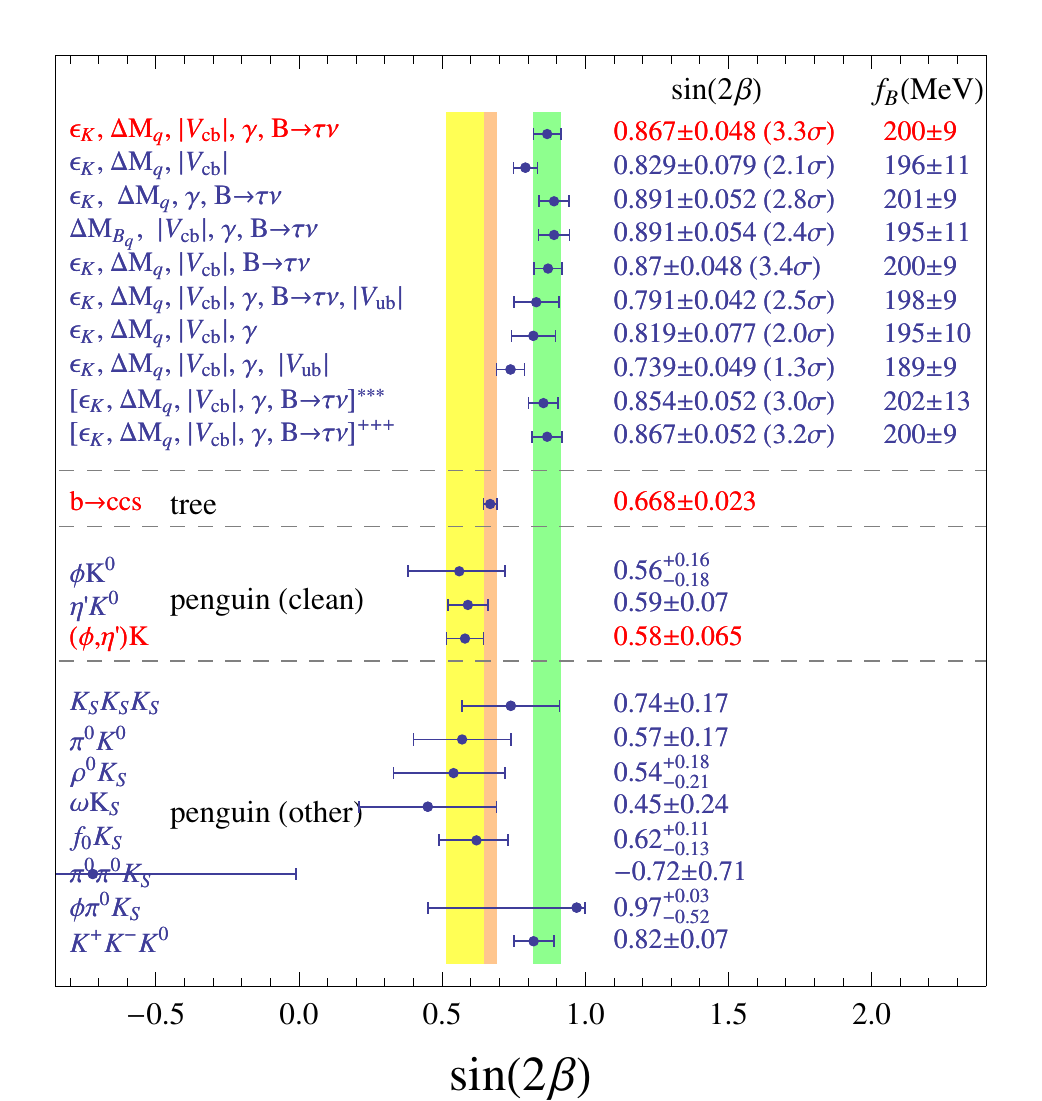}
\caption{Summary of $\sin(2\beta)$ determinations. The entry marked *** (ninth from the top) is obtained with lattice errors increased by 50\% over those given in Table~\ref{tab:utinputs} for each of the input quantities that we use and
the entry marked +++ (tenth from the top) corresponds to adding an hadronic uncertainty $\delta \Delta S_{\psi K} = 0.021$ to the relation between $\sin (2\beta)$ and $S_{\psi K}$. See the text for further explanations. \label{fig:tabsin2beta}}
\end{center}
\end{figure}

This conclusion is corroborated by the observation that even without using $B\to\tau\nu$ at all, and using as input only $\epsilon_K$, $\Delta M_{B_s}/\Delta M_{B_d}$ and $|V_{cb}|$ (see Fig.~\ref{fig:tabsin2beta}), the predicted value of $\sin(2\beta)$ deviates by $2.1\sigma$ from its measurement (in this case we find $\sin(2\beta)^{\rm fit} = 0.829 \pm 0.079$). Thus,  possible new physics in $B\to\tau\nu$ can alleviate but not remove completely the tension in the fit.

We recall that the fit above is actually  the simple fit we had reported some time ago (now with updated lattice inputs) with its resulting  $\approx$ 2 $\sigma$ deviation~\cite{Lunghi:2008aa}. This fit is somewhat special as primarily one is only using $\Delta F = 2$ box graphs from $\epsilon_K$ and $\Delta M_{B_s}/\Delta M_{B_d}$ in conjunction with lattice inputs for $B_K$ and the SU(3) breaking ratio $\xi$. The experimental input from box graphs is clearly short-distance dominated and for the lattice these two inputs are particularly simple to calculate as the relevant 4-quark operators have no mixing with lower dimensional operators and also require no momentum injection. The prospects for further improvements in these calculations are high and the method should continue to provide an accurate and clean ``prediction" for $\sin (2 \beta)$ in the SM. So even if the current tensions get resolved, this type of fit should remain a viable way to test the SM as lattice calculations and experimental inputs continue to improve.
\\

\begin{figure}[t]
\begin{center}
\includegraphics[width=0.95 \linewidth]{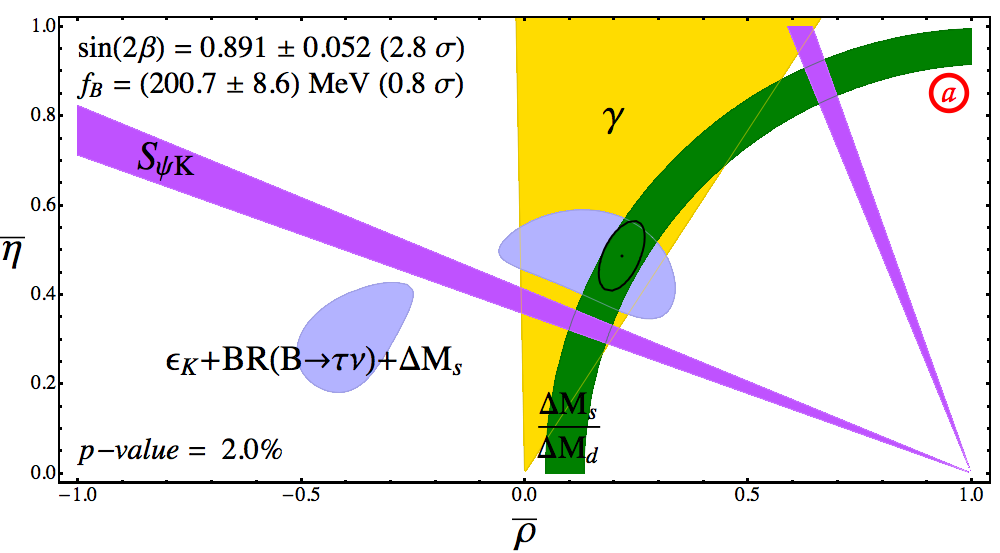}
\includegraphics[width=0.95 \linewidth]{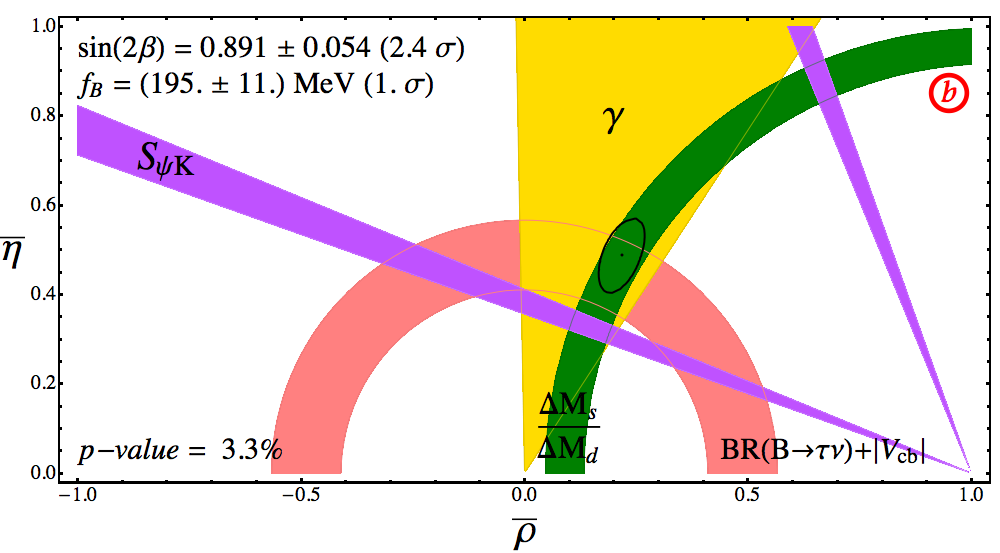}
\caption{Unitarity triangle fit without semileptonic decays (upper panel) and without use of $K$ mixing (lower panel). \label{extrafits}}
\end{center}
\end{figure}
\noindent {\bf Roles of $V_{cb}$, $\varepsilon_K$, $V_{ub}$ and of hadronic uncertainties.} 
The fit described above does use $V_{cb}$ where again the inclusive and exclusive methods differ mildly (about 1.7$\sigma$). Of greater concern here is that $\epsilon_K$ scales as $|V_{cb}|^4$ and therefore is very sensitive to the error on $V_{cb}$. We address this in two ways. First in Fig.~\ref{extrafits}a we study a fit wherein no semi-leptonic input from $b \to c$ or $b \to u$ is being used. Instead, in this fit ${\rm BR} (B \to \tau \nu)$ and $\Delta M_{B_s}$ along with $\epsilon_K$, $\Delta M_{B_s}/\Delta M_{B_d}$ and $\gamma$ are used. Interestingly this fit gives
\begin{align}
\sin(2\beta)^{\rm fit} &= 0.891 \pm 0.052 \; , \label{no_vcb} \\
f_B^{\rm fit} &= (200.7 \pm 8.6)\; {\rm MeV} 
\end{align}
Thus, once again, $\sin (2 \beta)$ is off by $2.8\sigma$ whereas $f_B$ is in very good agreement with directly measured value which we again take to mean that the {\it bulk} of the discrepancy is in $\sin (2 \beta)$ rather than in $B \to \tau \nu$ or in $V_{cb}$.

Next we investigate the role of $\epsilon_K$. In Fig.~\ref{extrafits}b  we show a fit where only input from B-physics, namely $\Delta M_{B_s}/\Delta M_{B_d}$, $\Delta M_{B_s}$, $\gamma$, $|V_{cb}|$ and ${\rm BR} (B \to \tau \nu)$ are used. This fit yields,
\begin{align}
\sin(2\beta)^{\rm fit} &= 0.891 \pm 0.054 \; , \label{no_epsilonk} \\
f_B^{\rm fit} &= (195 \pm 11)\; {\rm MeV} 
\end{align}
Thus, $\sin(2\beta)^{\rm fit}$ is off by $\approx 2.4 \sigma$ and again $f_B^{\rm fit}$ is in good agreement with its direct determination. We are, therefore, led to conclude that the role of $\epsilon_K$ in the discrepancy is subdominant and that the bulk of the new physics contribution is likely to be in B--physics. As before, the fact that the fitted value of $f_B$ is in good agreement with its direct determination seems to suggest that the input ${\rm BR}(B\to \tau \nu)$ is most likely not in any large conflict with the SM, though, obviously we cannot rule out the possibility of it receiving a sub-dominant contribution from new physics.
\begin{figure}[t]
\begin{center}
\includegraphics[width=0.95 \linewidth]{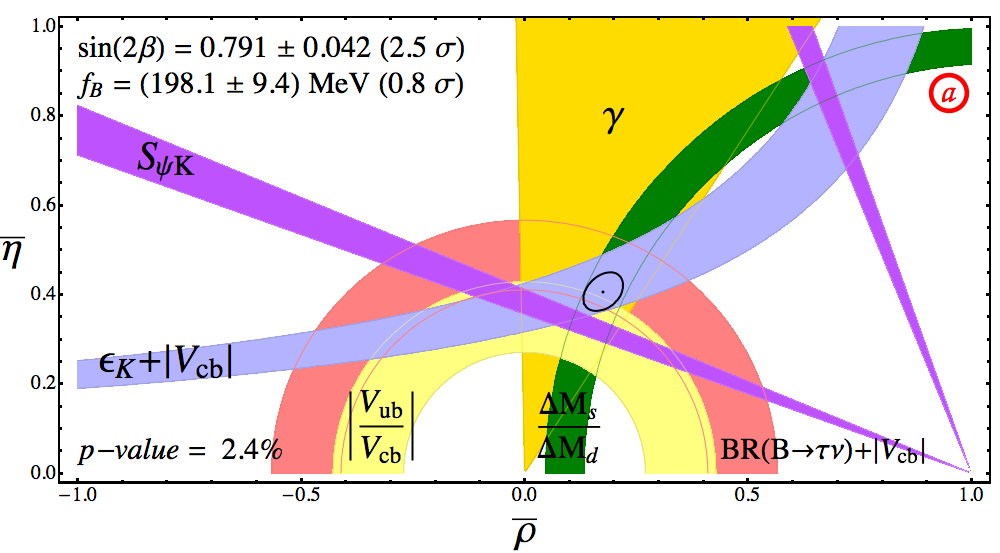}
\includegraphics[width=0.95 \linewidth]{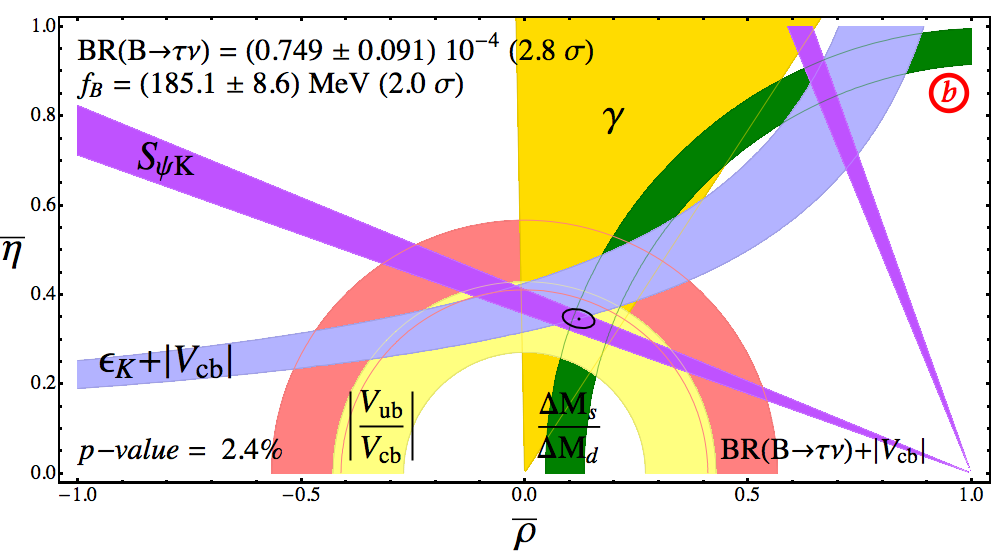}
\caption{Unitarity triangle fit with $V_{ub}$. See the caption in Fig.~\ref{fig:utfit}. \label{fig:utfit_vub}}
\end{center}
\end{figure}

Although we believe that we have been very careful in taking the input from lattice calculations and their  associated errors (see Table~\ref{tab:utinputs}), to gain further confidence we study the effect of increasing the total  error in each of the input quantity by 50\%, we find that qualitatively little change takes place from Eq.~(\ref{sin2betafit}):
\begin{align}
\sin(2\beta)^{\rm fit} &= 0.854 \pm 0.052 \; , \label{largererrors} \\
f_B^{\rm fit} &= (202 \pm 13)\; {\rm MeV} 
\end{align}
again we find a 3.0 $\sigma$ deviation in $\sin (2\beta)$ from the measured value and the fitted value of $f_B$ in very good agreement with the direct lattice determination strongly suggesting once again that the discrepancy with the CKM is rather serious.

So far we have stayed away from using $|V_{ub}|$ because of the large associated uncertainties with it. As another rough consistency check, let us mention that with the inclusion of $|V_{ub}|$ (resulting from combining inclusive and exclusive methods with an estimated error of about $15\%$ -- see Table~\ref{tab:utinputs}), the results  outlined above do not change qualitatively. The discrepancy between the fitted and measured values of $\sin (2 \beta)$ is mildly reduced to about 2.5$\sigma$ (see Fig.~\ref{fig:utfit_vub}a) while the fit result for $f_B$ is quite compatible with its direct lattice determination. In Fig.~\ref{fig:utfit_vub}b we investigate the alternate hypothesis of using S($\psi Ks$) as a SM input and find that even with the use of $|V_{ub}|$ the results remain essentially unchanged: both ${\rm BR}(B \to \tau \nu)$ and $f_B$ deviate appreciably from their respective direct determinations.

A compilation of all ten fits that we study for $\sin 2 \beta$ are shown in Fig.~\ref{fig:tabsin2beta}. Notice that there is only one case in here (8th from the top) where the discrepancy in $\sin 2 \beta$ is only $O(1\sigma)$. We believe this is 
primarily a reflection of the large ($\approx 14.4\%$) uncertainty with our combined $V_{ub}$ fit originating from the large disparity between inclusive and exclusive determinations; this is why we make only a limited use of $V_{ub}$ in our fits. 

\begin{figure}
\begin{center}
\includegraphics[width=0.95 \linewidth]{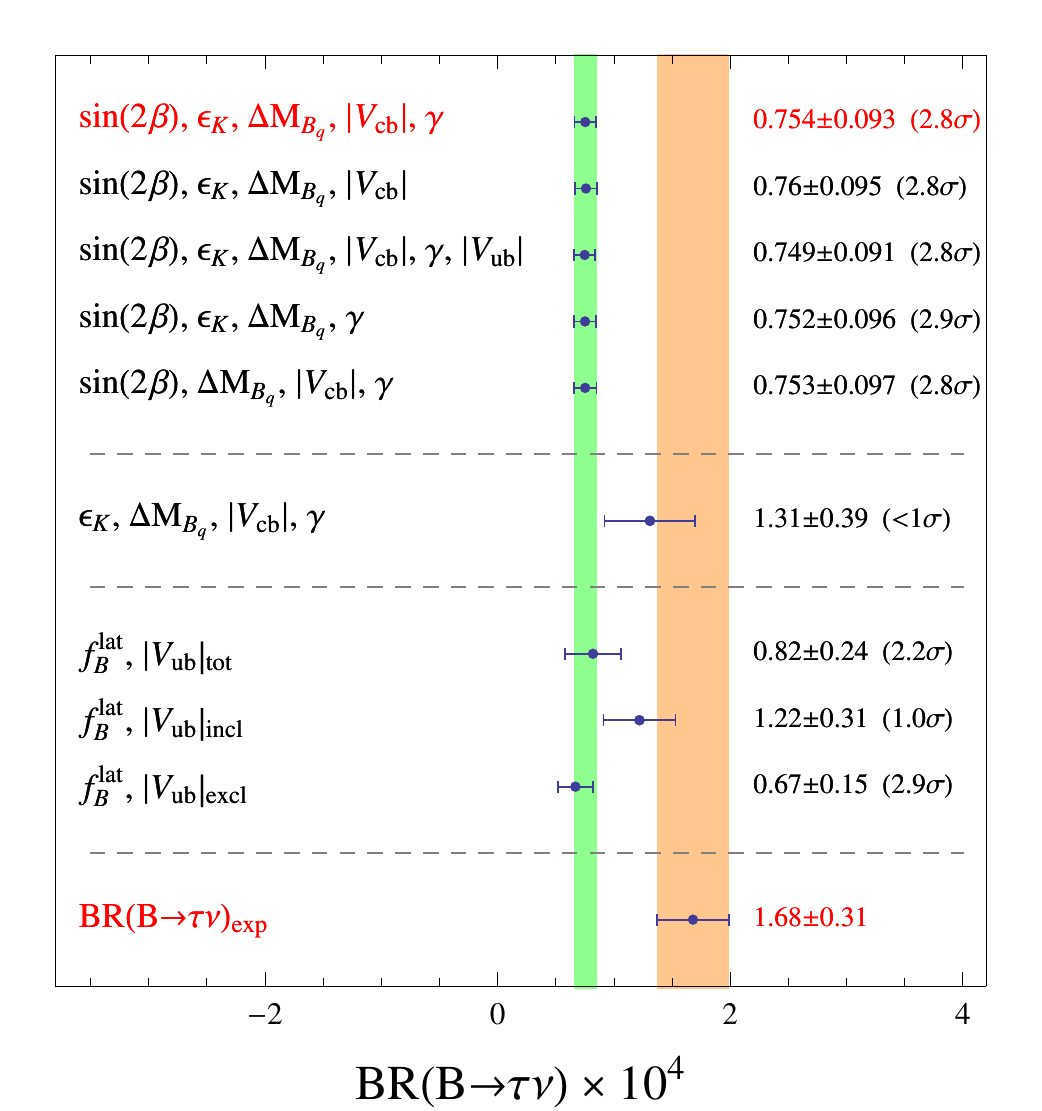}
\caption{Summary of  ${\rm BR} (B\to\tau\nu)$ determinations. \label{fig:tabbtn}}
\end{center}
\end{figure}

Now with regard to $B \to \tau \nu$, Fig.~\ref{fig:tabbtn} shows a summary of predictions versus the measured ${\rm  BR}$. Notice that whenever  the measured value of $\sin ( 2 \beta)$ is used as an  input, the predicted BR is $\approx 2.8 \sigma$ from the measured one. In the preceding discussion we have emphasized that this seems to us to be a consequence of new physics largely in $B$ mixings. This conclusion receives further strong support when we try determine the $B\to\tau\nu$ branching ratio without using $\sin 2 \beta$; indeed as shown in Fig.~\ref{fig:tabbtn} when we use $\epsilon_K$, $\Delta M_{B_q}$, $V_{cb}$ and $\gamma$ only the fitted value of ${\rm BR}(B \to \tau \nu)$ is in very good agreement with the measured value. 

In principle, of course the prediction for ${\rm  BR}(B \to \tau \nu)$ only needs the values of $f_B$ and of $V_{ub}$. Fixing now $f_B = 208 \pm 8~{\rm MeV}$ as  directly determined on the lattice (see Table~\ref{tab:utinputs}) we show the corresponding two predictions for the BR using separately the values of $V_{ub}$ determined in inclusive and exclusive decays. It is clear that the inclusive determination yields results that are within one $\sigma$ of experiment (see also Fig.~\ref{fig:utfit}); however with $V_{ub}$ from exclusive modes (that makes use of the semileptonic form factor as determined on the lattice), the BR deviates by $\approx 3 \sigma$ from experiment. This may be a hint that lattice based exclusive methods have some intrinsic difficulty or that the exclusive modes are sensitive to some new physics that the inclusive modes are insensitive to, {\it e.g.} right-handed currents~\cite{Crivellin:2009sd,Buras:2010pz}. In either case, this reasoning suggests that we  try using  the value of  $V_{ub}$ given by inclusive methods only in our fit for determining $\sin 2 \beta$. We then obtain results that are very compatible with the no--$V_{ub}$ case: the fitted value of $\sin(2\beta)$ deviates by $3.2\sigma$ from its direct determination (see Fig~\ref{fig:tabsin2beta}).

Finally, we address the possible presence of sizable hadronic uncertainties in the relation between $\sin(2\beta)$ and $S_{\psi K}$. Naive estimates of the impact that the CKM suppressed $u$-penguin amplitude, which is causing the ``pollution",  has on $S(B\to J/\psi K)$ point to a sub-percent effect. Quantitative studies based on QCD--factorization~\cite{Boos:2004xp}, perturbative QCD~\cite{Li:2006vq} and conservative model independent bounds on possible rescattering effects~\cite{Bander:1979px} corroborates~\cite{Gronau:2008cc} the above mentioned naive expectations. An alternative approach based on use of flavor $SU(3)$ to connect $B\to J/\psi K$ and $B\to J/\psi \pi$ hadronic matrix elements has been proposed in Refs.~\cite{Ciuchini:2005mg, Faller:2008zc}. This method is based on the observation that, up to $SU(3)$ corrections, the tree and penguin matrix elements appearing in $B\to J/\psi K$ and $B\to J/\psi \pi$ decays are identical. Since the penguin topology in $B\to J/\psi \pi$ is not CKM suppressed with the respect to the corresponding tree amplitude, time--dependent CP asymmetries in $B\to J/\psi \pi$ are highly sensitive to effects that affect the $J/\psi K$ mode at the percent level. Unfortunately, data on $B\to J/\psi \pi$ CP asymmetries is not precise enough to offer a measurement of the penguin to tree ratio and phase and we do not reliably know how large should be the SU(3) breaking.  The upper limits are presently more than  an order of magnitude above all the direct estimates. For these reasons we believe that, presently, it is not useful to adopt  $B\to J/\psi \pi$ decays as a sole handle on the size of penguin pollution in $B\to J/\psi K$. For completion we mention that even adopting the estimate of Ref.~\cite{Ciuchini:2010} for penguin pollution in the extraction of $\sin(2\beta)$, i.e. $\delta \Delta S = 0.021$, Eq.~(\ref{sin2betafit}) deviates from the measurement at the $3.2\sigma$ level.

\noindent {\bf Summary, perspective \& outlook.} 
In this paper we have mainly emphasized that our analysis strongly suggests that the SM predicted value of $\sin ( 2 \beta)$ is  around 0.85 whereas the value measured experimentally via the gold plated $\psi K_s$ mode is around 0.66 constituting a deviation of about 3 $\sigma$ from the SM (see Fig.~\ref{fig:tabsin2beta}). To put this result in a broader perspective let us now recall that in fact in the SM $\sin (2 \beta)$ can also be measured via the penguin dominated modes (see Fig.~\ref{fig:tabsin2beta})~\cite{Grossman:1996ke,Fleischer:1996bv,Grossman:1997gr,London:1997zk}. Unfortunately these modes suffer from a potentially large penguin pollution, though there are good reasons to believe that the $\eta^{\prime} K_s$, $\phi K_s$ and 3 $K_s$ modes are rather clean~\cite{Cheng:2005bg,Cheng:2005ug,Beneke:2005pu} wherein the deviations from $\sin 2 \beta$ are expected to be only O(few \%). The striking aspect of these three clean modes as well as many others penguin dominated modes (see Fig.~\ref{fig:tabsin2beta}) is that the central values of almost all of them tends to be even smaller than the value (0.66), measured in $\psi K_s$, and consequently tend to exhibit even a larger deviation from the SM prediction of around 0.85. Thus, seen  in the light of our analysis, the deviation in these penguin modes suggests the presence of new CP-violating physics not just in $B$-mixing but also in $b \to s$ penguin transitions.

Moreover, the large difference ($\approx (14.4 \pm 2.9) \%$)~\cite{PDG_10} in the direct CP asymmetry  measured in $B^0  \to K^+ \pi^-$  versus that in $B^+ \to K^+ \pi^0$ provides another hint that $b \to s$ penguin transitions may be receiving the contribution from a beyond the SM source of CP-violation. To briefly recapitulate, in the SM one naively expects this difference to be vanishingly small and careful estimates based on QCD factorization ideas suggest that it is very difficult to get a difference much larger than $(2.2 \pm 2.4) \%$~\cite{Lunghi:2009sm}.

Of course, if $b \to s$ penguin transitions ($\Delta Flavor =1$)  are receiving contributions from new physics, then it is quite unnatural  for $B_s$ mixing amplitudes ($\Delta Flavor =2$) to remain unaffected. Therefore, this reasoning suggests that we should expect non-vanishing CP asymmetries in $B_s \to \psi \phi$ as well as a non-vanishing  di-lepton asymmetry in $B_s \to X_s l \nu$.  As is well known,  at Fermilab, in the past couple of years CDF and D0 experiments have been studying CP asymmetry in $B_s \to \psi \phi$. The latest analysis with about 6.1 ${\rm fb}^{-1}$ luminosity  at D0 shows a deviation from the SM of about 2 $\sigma$~\cite{RK_ICHEP10}, which is a mild increase from their previous result ($\approx 1.8 \sigma$) with 2.8 $fb^{-1}$ whereas at CDF the results with corresponding increase in luminosity has shown a downward shift in the deviation from the SM to  $\approx 0.8 \sigma$ from the earlier $\approx 1.8 \sigma$ deviation~\cite{DT_BF10}. Thus, the Fermilab  findings in $B_s \to \psi \phi$ are not yet clear. 

Another interesting and potentially very important development with regard to non-standard CP in $B_s$ is that in the last several months D0 has announced the observation of a large dimuon asymmetry in B-decays amounting to a deviation of ($\approx 3.2 \sigma$) from the miniscule asymmetry predicted in the SM~\cite{Abazov:2010hv,Abazov:2010hj}. They attribute this largely to originate from $B_s$ mixing. While this is a very exciting development, their experimental analysis is extremely challenging and a confirmation is highly desirable before their findings can be safely assumed.

Be that as it may, we reiterate that our analysis  suggests that the deviation from the SM in $\sin (2 \beta)$ is difficult to reconcile with errors in the inputs from the  lattice  that we use,  and strongly suggests the presence of a non-standard source of CP violation largely in $B$/$B_s$ mixings, thereby predicting that non-standard signals of CP violation in $S (B_d \to \eta^{\prime} K_s, \phi K_s, 3 K_s etc. )$ as well as in $S(B_s \to \psi \phi)$, and  the semileptonic  and di-lepton asymmetries  in $B_s$, and possibly also in $B_d$, decays will persist and survive further scrutiny in experiments at the intensity frontiers such as Fermilab (CDF, D0), LHCb and the Super-B factories. Lastly, the fact that our analysis rules out the presence of large new physics in kaon mixing has very important  repercussions for the  mass scale of the underlying new physics contributing to  these anomalies in $B$, $B_s$ decays: model independent analysis then  imply that the relevant mass scale of the  new physics is necessarily relatively low, {\it i.e.} below O(2 TeV) and may, in principle, be as low as a few hundred GeVs~\cite{Lunghi:2009sm,no_LR}. Thus, collider experiments at the high energy frontier at LHC and possibly even at Fermilab should see direct signals of the  underlying degrees of freedom appearing in any relevant beyond the Standard Model scenario. 

\begin{acknowledgments}
We want to thank Elvira Gamiz and Ruth Van de Water for discussions.
This research was supported in part by the U.S. DOE contract No.DE-AC02-98CH10886(BNL).
\end{acknowledgments}


\begin{thebibliography}{99}
\bibitem{no_alpha}
In contrast to $\gamma$, $\alpha$ is not used as an input since it receives appreciable contribution from  penguin amplitudes which are sensitive to new physics.
\bibitem{no_gamma}
Note that when $\gamma$ is not used as an input and only $\epsilon_K$, $\Delta M_s/\Delta M_d$ and ${\rm BR} (B \to \tau \nu)$ are used, the deviation of the fitted $\sin (2 \beta)$ from the measured one increases mildly to 3.4$\sigma$, see Fig.~\ref{fig:tabsin2beta}.
\end{thebibliography}
\end{document}